\documentclass[epj,a4paper]{svjour}
\usepackage[T1]{fontenc}
\usepackage[latin1]{inputenc}
\usepackage{amsmath}
\usepackage{graphics}

\makeatletter

\providecommand{\LyX}{L\kern-.1667em\lower.25em\hbox{Y}\kern-.125emX\@}
\newcommand{\noun}[1]{\textsc{#1}}
\let\SF@@footnote\footnote
\def\footnote{\ifx\protect\@typeset@protect
    \expandafter\SF@@footnote
  \else
    \expandafter\SF@gobble@opt
  \fi
}
\expandafter\def\csname SF@gobble@opt \endcsname{\@ifnextchar[%]
  \SF@gobble@twobracket
  \@gobble
}
\edef\SF@gobble@opt{\noexpand\protect
  \expandafter\noexpand\csname SF@gobble@opt \endcsname}
\def\SF@gobble@twobracket[#1]#2{}

\usepackage[squaren]{SIunits}
\usepackage{units}
\usepackage{amssymb}
\usepackage{mathrsfs}
\usepackage{url}
\newcommand{\oper}[1]{\boldsymbol{\mathsf{#1}}}
\makeatother

\begin{document}

\title{Quantum mechanical ab-initio simulation of the electron screening effect in
metal deuteride crystals}

\author{A. Huke\inst{1} \and K. Czerski\inst{1,2} \and S. M. Chun\inst{1}\thanks{deceased} \and A. Biller\inst{1} \and P. Heide\inst{1}}
\institute{Institut für Optik und Atomare Physik,
Technische Universität Berlin, Hardenbergstr. 36, 10623 Berlin, Germany \and
Institute of Physics, University of Szczecin, Szczecin, Poland}
\mail{huke@physik.TU-Berlin.DE, Armin.Huke@web.de}
\date{}

\abstract{
In antecedent experiments the electron screening energies of the d+d reactions
in metallic environments have been determined to be enhanced by an order of
magnitude in comparison to the case of gaseous deuterium targets. The analytical
models describing averaged material properties have not been able to explain
the experimental results so far. Therefore, a first effort has been undertaken
to simulate the dynamics of reacting deuterons in a metallic lattice by means
of an ab-initio Hartree-Fock calculation of the total electrostatic force between
the lattice and the successively approaching deuterons via path integration.
The calculations have been performed for Li and Ta, clearly showing a migration
of electrons from host metallic to the deuterium atoms. However, in order to
avoid more of the necessary simplifications in the model the utilization of
a massive parallel supercomputer would be required.
}
\maketitle

\section{Introduction}

Following our first observation \cite{volos98,europhys01,dis} of the grossly
enhanced screening effect for the d+d reactions in metals resulting in an exponential-like
increase of the astrophysical S-factors towards lower energies other groups
reproduced our results \cite{yuki98,kasagi02,rolfs02,rolfs02b,rolfs03,rolfs04}.
The measured screening energies are one order of magnitude larger than for D\( _{2} \)-gas
targets \cite{greife95}. However, particular care is required for the interpretation
of the experimental data because the special physico-chemical properties of
the hydrogen compounds and the beam induced chemical reactions at the target
heavily influence the obtained screening energy results \cite{dis,nimb06}.
Originally, the screening energy was extracted using the simple model of \cite{assenbaum87}
treating it as an shift of the kinetic energy. Thus, the screening energy is
merely a parameter which describes the modification of the Coulomb barrier and
not a real energy gain for the projectile. In a microscopic view it is universally
valid that the screening effect depends on the impact of the electronic configuration
of the environment on the Coulomb barrier of the entrance channel only, i.e.~the
pure Coulomb energy is simply modified by a Yukawa factor \( W(r)=\frac{1}{4\pi \varepsilon _{0}}\frac{Z_{p}Z_{t}e^{2}}{r}e^{-\frac{r}{\lambda _{\mathrm{A}}}} \)with
\( \lambda _{\mathrm{A}} \) being the screening length. As such the inferred
screening energy is merely the second term in a Taylor-expansion of \( W(r) \),
i.e.~\( U_{e}=\frac{1}{4\pi \varepsilon _{0}}\frac{Z_{p}Z_{t}e^{2}}{\lambda _{\mathrm{A}}} \),
and a coarse mathematical parametrization in the simple model.

The experimental screening energy values could not be described by theoretical
approaches to this extent. In \cite{europhys04,NPAII06b} we presented an analytical
model based on the dielectric function theory for the description of the screening
effect. Its results, however, are below the measured screening energies by a
factor of 2 though better than other approaches cited therein, e.g.~\cite{ichimaru93}\footnote{%
The Debye-Hückel model proposed in \cite{rolfs03} is utterly inapplicable for
the frozen electron gas in condensed matter \cite{NPAII06b,prc07}.
}. Therefore, a first effort was undertaken to simulate the pre-reaction impact
with an ab-initio quantum mechanical Hartree-Fock calculation which is able
to consider the actual crystal structure while the analytical model only operates
on averaged material properties. The concrete form of the electron density distribution
around the nucleus indeed influences the screening energy to a great extent
as was shown in \cite{shoppa96} on a D\( _{2} \) molecule with a time dependent
Hartree-Fock calculation. Due to the many electrons involved in the simulation
of the crystal the time dependency can not be retained on a workstation class
computer. The model is described in the next section permitting the assessment
of the necessary simplifications.

\section{Mathematical Model}

The mathematical model, the simulation is based on, results from a hierarchy
of simplifications in the theoretical description of the physical reality leading
to more or less severe deviations in the obtained values of the observables.

The originator for it is the non-relativistic spin independent Schrödinger equation

\begin{equation}
\label{eq:tdschroedinger}
i\hbar \frac{\partial }{\partial t}\Psi =\oper {H}\Psi 
\end{equation}
 with the Hamilton operator for the crystal molecule 
\begin{eqnarray}
\oper {H} & = & -\sum ^{K}_{a}\frac{\hbar ^{2}}{2M_{a}}\boldsymbol {\Delta }_{a}+\frac{e^{2}}{4\pi \varepsilon _{0}}\sum ^{K}_{a<b}\frac{Z_{a}Z_{b}}{R_{ab}}\label{eq:HoperNuc} \\
 &  & -\sum ^{N}_{i}\frac{\hbar ^{2}}{2m_{e}}\boldsymbol {\Delta }_{i}-\frac{e^{2}}{4\pi \varepsilon _{0}}\sum _{i}^{N}\sum _{a}^{K}\frac{Z_{a}}{r_{ai}}\label{eq:HoperElec} \\
 &  & +\frac{e^{2}}{4\pi \varepsilon _{0}}\sum ^{N}_{i<j}\frac{1}{r_{ij}}\label{eq:HoperElecEX} 
\end{eqnarray}
 containing both kinetic and potential operators for the \( K \) nuclei (\ref{eq:HoperNuc})
and the \( N \) electrons (\ref{eq:HoperElec}, \ref{eq:HoperElecEX})\footnote{%
Notations according to \cite{schmidtke94}.
}. Therein \( Z_{a} \) is the charge and \( M_{a} \) the mass of the nucleus
\( a \). The distances in the denominators of the three potential operators
are the absolute values of the vector differences of the respectively indexed
particles. Inherent to this ansatz the internal degrees of freedom of the nuclei
are not taken into account. But the screening forces of the electrons act at
distances far beyond the range of the nuclear forces. The connection to the
nucleus takes place via the calculated screening energy \( U_{e} \) in the
cross-section. Just as little any spin dependent effects are considered. So
polarization effects as observed in \cite{NPAII06a} cannot be described. Since
the impact proceeds at \kilo\electronvolt-energies the inclusion of relativistic
effects is not required. Furthermore, the velocity of the nuclei is small in
comparison to the electrons\footnote{%
The electrons of the inner shells of heavy atoms attain relativistic velocities.
This is commonly considered by corrections.
}. Hence, the Born-Oppenheimer approximation \cite{born27} can be applied which
leads eventually to a drop out of the nuclear operators in (\ref{eq:HoperNuc}),
leaving (\ref{eq:HoperElec}, \ref{eq:HoperElecEX}). The electron wave functions
become then continuous functions of the nuclear coordinates. The movement of
the projectiles is treated classically owing to the acting forces. The now purely
electronic Schrödinger equation can nevertheless only be solved approximately,
which is done based on the Ritz theorem by minimization of the energy functional
\( E\left[ \Psi \right] =\left( \Psi ,\oper {H}\Psi \right)  \) with \( \Psi  \)
from a restricted function space \( \mathcal{F}' \) under control parameters.
This is performed with the self-consistent field method by Hartree and Fock
(HF). The electronic Hamilton operator is splitted up in a single electron operator
(\ref{eq:HoperElec}) and a part containing the interaction potential (\ref{eq:HoperElecEX}).
The single electron problem has likewise solution functions \( \varphi \left( \vec{r},\sigma \right) :=\psi \left( \vec{r}\right) \chi \left( \sigma \right)  \)
with the space function \( \psi  \) and the spin function \( \chi  \). The
solution of the multi-electron problem is constructed with the anti-symmetrization
operator as a Slater determinant \( \Psi =\oper {A}\, \varphi _{\mathrm{a}}\left( 1\right) \cdot \varphi _{\mathrm{b}}\left( 2\right) \cdot \ldots \cdot \varphi _{\mathrm{n}}\left( N\right)  \)
in accord with the Pauli principle. The variation of the energy functional yields
the Hartree-Fock equations \cite{hartree57} 
\begin{eqnarray}
\oper {H}_{\mathrm{HF}}\psi _{\mathrm{a}}\left( i\right)  & = & -\frac{1}{2}\boldsymbol {\Delta }_{i}\psi _{\mathrm{a}}\left( i\right) -\sum _{a}^{K}\frac{Z_{a}}{r_{ai}}\psi _{\mathrm{a}}\left( i\right) \nonumber \\
 &  & +\sum _{\mathrm{b}}^{n/2}\left( 2\oper {J}_{\mathrm{b}}-\oper {K}_{\mathrm{b}}\right) \psi _{\mathrm{a}}\left( i\right) \nonumber \\
 & = & e_{\mathrm{a}}\psi _{\mathrm{a}}\left( i\right) \label{eq:HF} 
\end{eqnarray}
with the Coulomb operator \( \oper {J}_{\mathrm{b}} \) and the exchange operator
\( \oper {K}_{\mathrm{b}} \) defined as 
\begin{eqnarray}
\oper {J}_{\mathrm{b}}\psi _{\mathrm{a}}\left( i\right)  & = & \left( \psi _{\mathrm{b}}\left( j\right) ,\frac{1}{r_{ij}}\psi _{\mathrm{b}}\left( j\right) \right) \psi _{\mathrm{a}}\left( i\right) \nonumber \label{m:screen.JCoul} \\
\oper {K}_{\mathrm{b}}\psi _{\mathrm{a}}\left( i\right)  & = & \left( \psi _{\mathrm{b}}\left( j\right) ,\frac{1}{r_{ij}}\psi _{\mathrm{a}}\left( j\right) \right) \psi _{\mathrm{b}}\left( i\right) \label{m:screen.JK-Oper} 
\end{eqnarray}
 They have the form of an eigenvalue equation for \( \psi _{\mathrm{a}} \)
with the orbital energy \( e_{\mathrm{a}} \), though they are coupled by the
operators \( \oper {J}_{\mathrm{b}} \) and \( \oper {K}_{\mathrm{b}} \). Indeed,
they are a system of coupled integro-differential equations which can only be
solved iteratively. To simplify matters the case of the restricted HF method
is described where all \( n/2 \) orbitals are occupied with electrons. Otherwise
the unrestricted HF method has to be used \cite{levine91,szabo89}. The HF method
furnishes merely a stationary solution so far. A time development can be achieved
with the unitary time propagation operator \( \psi \left( t\right) =\oper {U}\left( t,t_{0}\right) \psi \left( t_{0}\right)  \)
which is given by \( \oper {U}\left( t,t_{0}\right) =e^{-i\left( t-t_{0}\right) \oper {H}} \)
if the Hamilton operator is not explicitly time dependent \cite{messiah91a}.
Caley's form provides a unitary approximation for this equation as a calculable
time discretization resulting in the time dependent Hartree Fock method using
\( \oper {H}_{\mathrm{HF}} \) \cite{goldberg67,galbraith84}. Such has been
carried out by \cite{shoppa93,shoppa96} implementing the discretization of
the electron wavefunctions on a space lattice.

The HF method cannot include immediate electron-electron interaction\footnote{%
The operators \( \oper {J}_{\mathrm{b}} \) and \( \oper {K}_{\mathrm{b}} \)
in (\ref{m:screen.JK-Oper}) merely provide an averaged effective electron potential
in the iteration due to the integration in the scalar product, e.g.~\cite{schmidtke94}.
}. For electrons of equal spin the main correlation effects are incorporated
in the Slater determinant. But the motion of electrons of opposite spin remains
uncorrelated. To address this, several procedures have been developed which
either start from the HF solution functions applying corrections or modify the
HF method \cite{szabo89,carsky80,zuelicke73,bartlett89}. They are based on
the configuration interaction method (CI) or the M{\o}ller-Plesset perturbation
theory (MPn). The CI method uses the HF orbital function and adjoins further
wherein occupied orbital functions are replaced by unoccupied virtual orbitals
successively thus forming a series \cite{levine91}. The linear coefficients
are determined by minimization of the energy functional. This series is infinite
and would deliver an exact solution for the Schrödinger equation (\ref{eq:HoperElec},\ref{eq:HoperElecEX})
in the limit. In practice, the series is aborted after the first to fourth term
with few virtual orbitals in the individual terms. The MPn method is essentially
an implementation of the common quantum mechanical perturbation theory \cite{moller34}.
It is usually aborted after the second to eighth order. Another approach is
the density functional theory (DFT) \cite{parr89,labanowski91}. Following the
theorem of Hohenberg and Kohn a unique functional exists which provides an exact
solution of (\ref{eq:HoperElec},\ref{eq:HoperElecEX}) \cite{hohenberg64,kohn65}.
But there is no algorithm that can generate this functional so one needs to
settle on heuristic ones and therefore the DFT is no ab-initio method in the
end \cite{ziegler91,vosko80,becke88,lee88,miehlich89,becke88b,johnson93}. The
DFT modifies the HF equations with an exchange interaction functional, which
is to be supplied.

The just described configuration interaction procedures can increase the computational
costs by 1 to 3 orders of magnitude relative to the plain HF method \cite[chapter 6]{foresman96}.
On the other hand the implementation of the orbital functions on space lattices
is very cost-intensive and only feasible for few electrons. Hence the analytic
HF method of Roothaan and Hall is used for larger systems \cite{roothaan51,hall51,pople54}.
There the orbital functions are developed as a linear combination \( \psi _{\mathrm{a}}=\sum _{n}^{k}c_{n\mathrm{a}}u_{n} \)
of a set of analytic functions \( \left\{ u_{n}\right\} ^{k}_{1}\in \mathrm{L}^{2}\left( \mathbb {R}^{3}\right)  \)
which are the basis of the finite dimensional product space \( \mathcal{F}'=\bigotimes ^{k}_{n}\mathcal{F}'_{n} \).
Thus the HF equations (\ref{eq:HF}) are transformed to a system of pseudo-linear
equations for the linear coefficients which are solved iteratively again. The
main computational costs now results from the integrations of the numerous scalar
products. The computationally most efficient basis functions for it are the
Cartesian Gauß type orbital functions \cite{preuss62} 
\begin{equation}
\label{eq:GTO}
u\left( \textrm{GTO}\right) =x^{i}y^{j}z^{k}e^{-\alpha r^{2}}\; ,\quad i,j,k\in \mathbb {N}_{0}
\end{equation}
 with the sum \( l=i+j+k \) yielding the angular momentum quantum number of
the orbital. Thereupon basis sets of increasing complexity are constructed being
geared to actual atomic orbitals \cite{szabo89}. The computer resource usage
scales with the total number of GTO basis functions as \( \mathcal{O}\left( n\right) \sim n^{4} \)
customarily spanning two orders of magnitude \cite[chapter 5]{foresman96}.
The common minimal basis set is STO-3G where the atomic orbitals are approached
by Slater type orbitals which contain the spherical harmonics and are constructed
from 3 GTO's with fixed coefficients, therefore called contracted \cite{hehre69,collins76}.
Split valence basis sets provide two or more sizes of basis functions for the
valence orbitals (e.g.~D95 \cite{dunning76}, LanL2DZ \cite{hay85,wadt85,hay85b},
3-21G \cite{blinkley80,gordon82,pietro82}, 6-31G \cite{ditchfield71,hehre72,hariharan74,gordon80,hariharan73},
6-311G). Polarized basis sets allow the atomic orbitals to change their shape
by providing orbitals with angular momentum higher than for the ground state
description, like p and d orbitals (here 6-31G(d,p) \cite{petersson88,petersson91}).
High angular momentum basis sets add more orbitals of p, d, and f type (here
6-311+G(3df,3pd) \cite{mclean80,krishnan80,wachters70,hay77,raghavachari89}).
The inclusion of diffuse functions adds larger size versions of s and p type
functions allowing the orbitals to cover larger regions of space which is advantageous
for charge transfers in ionic bonds (indicated by a '+'). Most of the basis
sets are only available for the lower periods of the periodic table. Only the
LanL2DZ covers heavier atoms where the inner shells are described by effective
core potentials including relativistic effects. The extra basis (+XB) adds diffuse
functions.

In consequence, it becomes clear that the simulation of the pre-reaction impact
inside the metal lattice involves so many electrons that the task has very high
computer resource requirements. For the given workstation (UltraSPARC), the
problem is actually oversized. So it is unevitable to drop the time dependence
and settle for the analytic HF method with restricted basis sets and few atoms,
possibly augmented by the DFT. The impact is then treated as quasi static. The
Coulomb adjusted relative electronic force between the projectile and target
nuclei is recorded and integrated along the trajectory \( T \) analogous to
\cite{shoppa93,shoppa96} yielding the molecular screening energy 
\begin{equation}
\label{m:screen.PfadInt}
U_{\mathrm{mol}}=\int\limits _{T}\vec{F}_{\mathrm{rel}}\cdot d\vec{r}
\end{equation}
 At the classical turning point the molecular screening energy is identified
with the screening energy \( U_{e}=-U_{\mathrm{mol}}\left( r_{cl}\right)  \).

So even for stationary orbital functions they are only feasible for relatively
small molecules or require super computer scale power.

\section{Calculations}

\begin{figure}
{\par\centering \resizebox*{0.7\columnwidth}{!}{\includegraphics{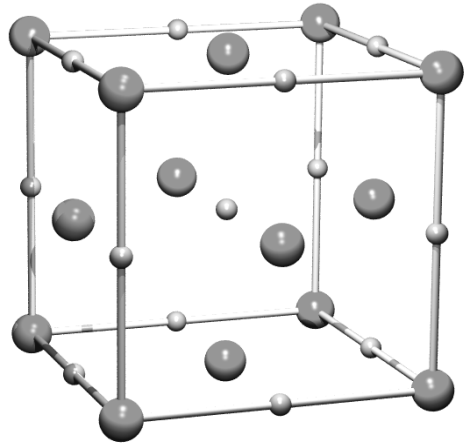}} \par}

\caption{\label{fig:KG-LiH}Cubic space centered crystal cell of LiH.}
\end{figure}

\begin{figure}
{\par\centering \resizebox*{0.7\columnwidth}{!}{\includegraphics{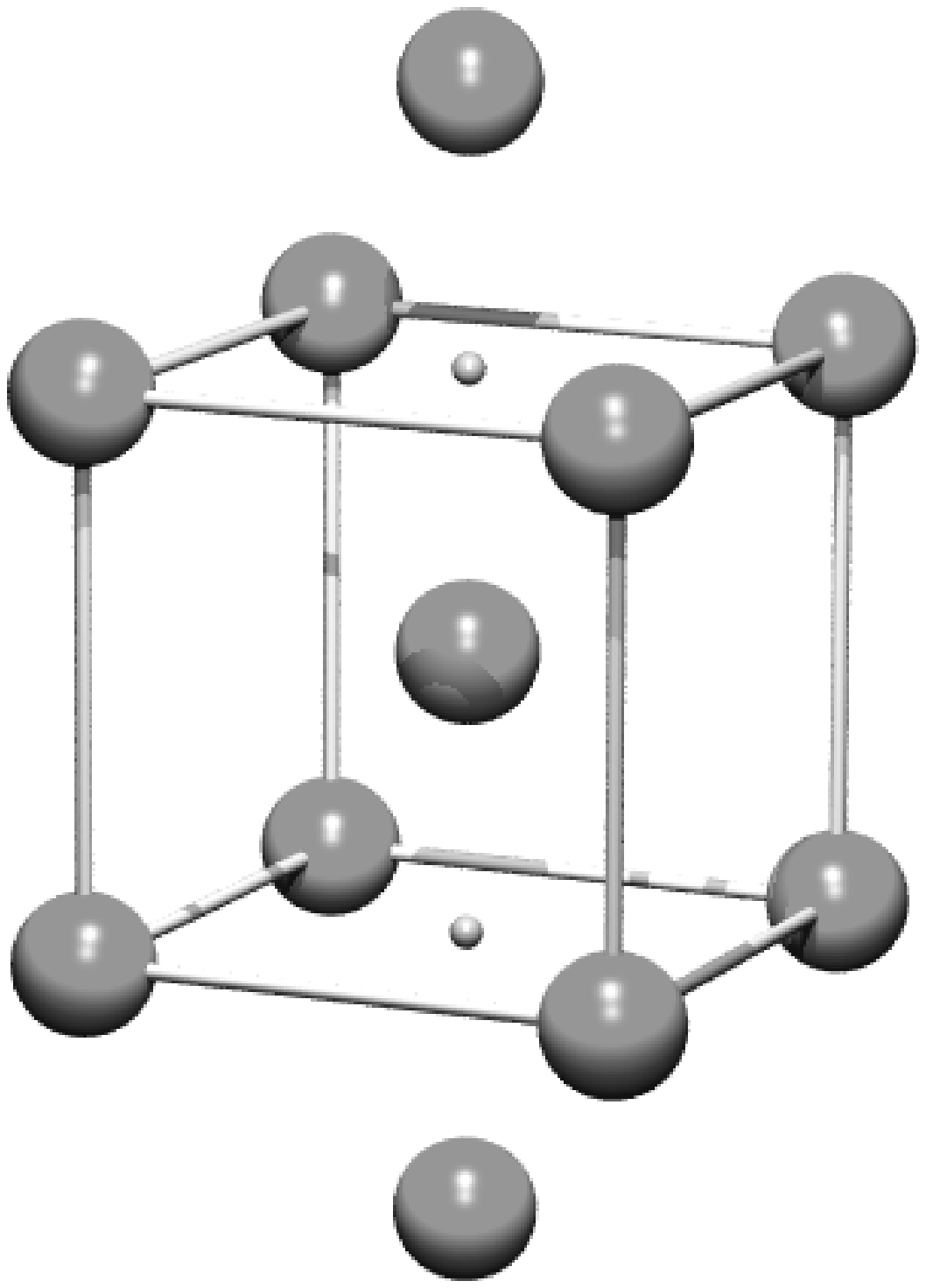}} \par}

\caption{\label{fig:KG-Ta2H}Tetragonal crystal cell of Ta\protect\( _{2}\protect \)H.}
\end{figure}

The here adopted crystal structure data are taken from \cite{crystal626}. LiH
has a cubic space centered structure with \( a=4.085\, \angstrom  \) \cite[III, a1]{crystal626}
fig.~\ref{fig:KG-LiH}. 
  Ta\( _{2} \)H has a tetragonal structure with \( a=3.38\, \angstrom  \),
\( b=3.41\, \angstrom  \) \cite[IV, j2]{crystal626} fig.~\ref{fig:KG-Ta2H}. 

\begin{figure}
{\par\centering \resizebox*{1\columnwidth}{!}{\includegraphics{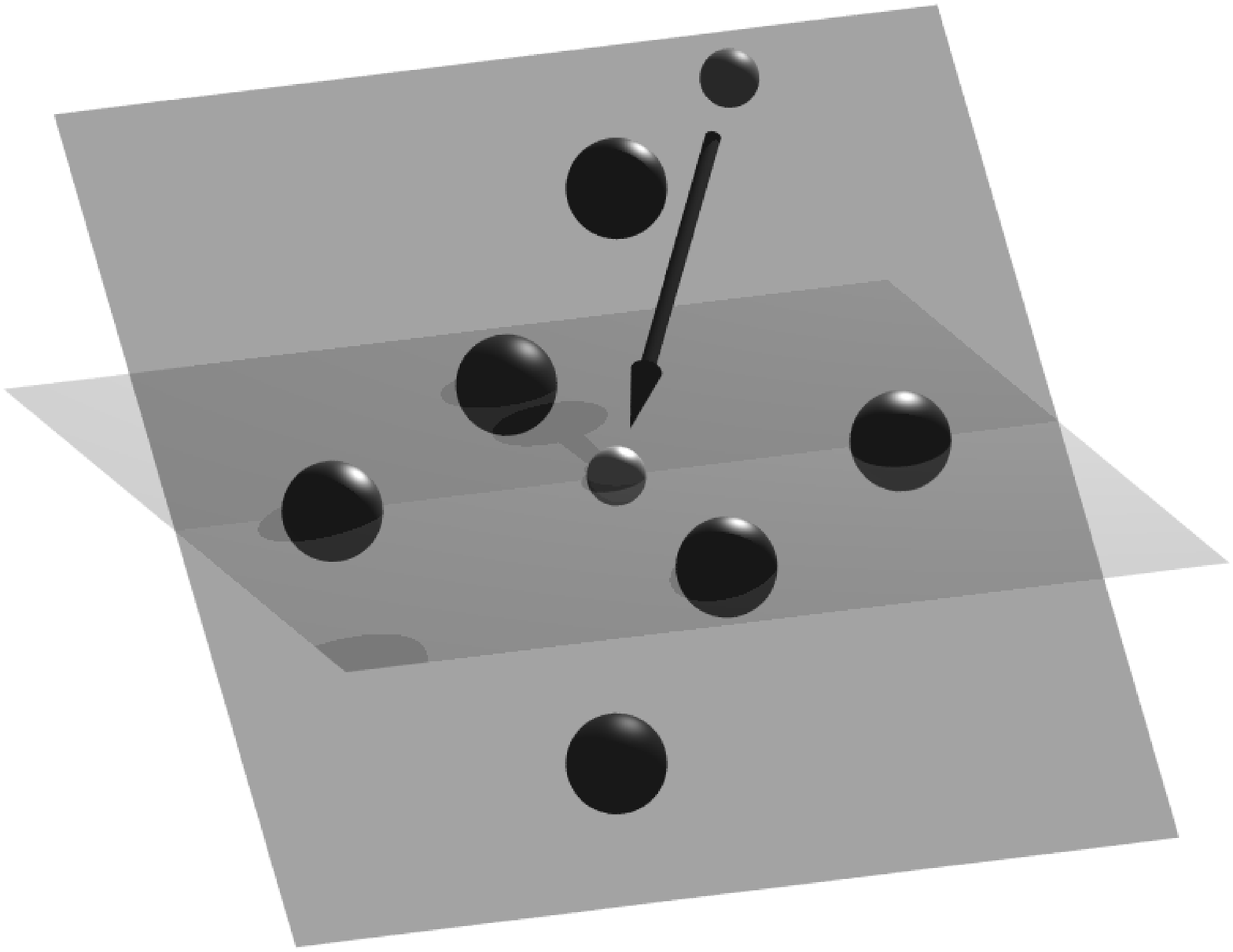}} \par}

\caption{\label{fig:Mol.lil-li1}LiH molecule (\texttt{lih-li1}) with projectile trajectory
and intersections.}
\end{figure}

\begin{figure*}
{\par\centering \resizebox*{0.9\textwidth}{!}{\includegraphics{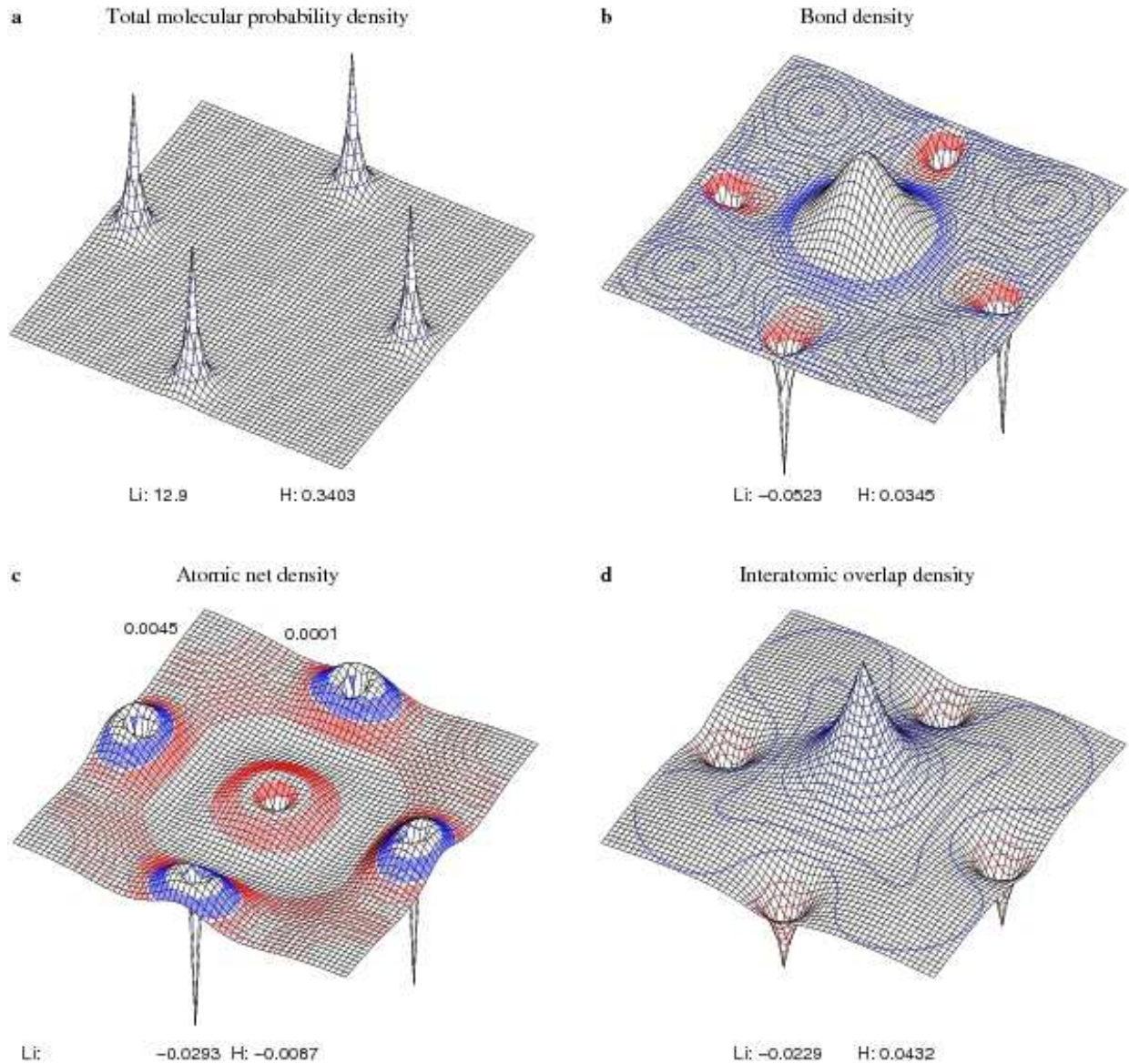}} \par}

\caption{\label{fig:DichteLil}Different electronic densities in the molecule \texttt{lih-li1}
with the basis 6-31G(d,p)}
\end{figure*}

The calculations were performed with different crystalline molecules, basis
sets and models for the electron exchange interaction employing the quantum
chemical package \noun{Gaussian} \cite{gaussian94}. One such molecule for
LiH is depicted in fig.~\ref{fig:Mol.lil-li1}. 
 The behaviour of the electrons in this molecule becomes clear by means of the
density plots in the intersection through the 4 Li atoms in fig.~\ref{fig:DichteLil}. 
 The total propability density of the electron wavefunction of the molecule
is plotted in fig.~\ref{fig:DichteLil}.a. The maximum values are numeralized
at the positions of the nuclei. The molecular wavefunction is not normalized
to unity but to the number of electrons, i.e.~\( \int _{\mathbb {R}^{3}}\Psi ^{*}\Psi \, d\vec{r}=n \).
So the probability density values can be larger than 1 and may allow for certain
statements on the electron number distribution. The electrons preferentially
sojourn at the three fold charged Li nuclei, of course. Hence, the total molecular
probability density is not very meaningful in order to observe the shift of
the electrons due to the chemical bond. Much more suitable is the bond density
(fig.~\ref{fig:DichteLil}.b) which is the difference between the total molecular
probability density and the probability density of the single atoms. The result
is a plot whose average value is zero and which shows the electron migration
when the bonds are formed. Now, it is visible that indeed an electron transfer
towards the central hydrogen atom at the expense of the Li atoms took place.
The density matrix of the bond density is composed of two parts: the atomic
net density (fig.~\ref{fig:DichteLil}.c) which results from setting the interatomic
overlap equal to zero. One recognizes where the electrons are deducted from
the atoms in order to form the bonds; from Li in a stronger proportion than
from H. The respective maximum values are marked above the plot and the minimum
values below. In the overlap density (fig.~\ref{fig:DichteLil}.d) the atomic
net density is set to zero and one recognizes the share of the molecular bond.
Due to the bond formation the hydrogen atom profits by gaining electron shares.
The ionic character of the bond becomes clear, since the overlap at the position
of the hydrogen nucleus becomes a maximum and not in between the atoms as in
the covalent bond.

The projectile trajectory is chosen so that the projectile does not close in
on other nuclei. According to \cite{shoppa96} on a nearby passage the spectating
nuclei exert a repulsive force on the projectile causing a reduction of the
screening energy, on average. On the other hand ions traversing crystal lattices
are deflected by the lattice atoms on trajectories inbetween them, known as
the channeling effect. This means here that the crystal lattice of the metal
atoms focuses the incoming deuterons on the interlaced lattice of the deuterons
in the metal hydride \cite{nimb02} counteracting the reduction of the screening
energy in the D\( _{2} \) molecule of \cite{shoppa96}.

\begin{figure*}
{\par\centering \resizebox*{0.9\textwidth}{!}{\includegraphics{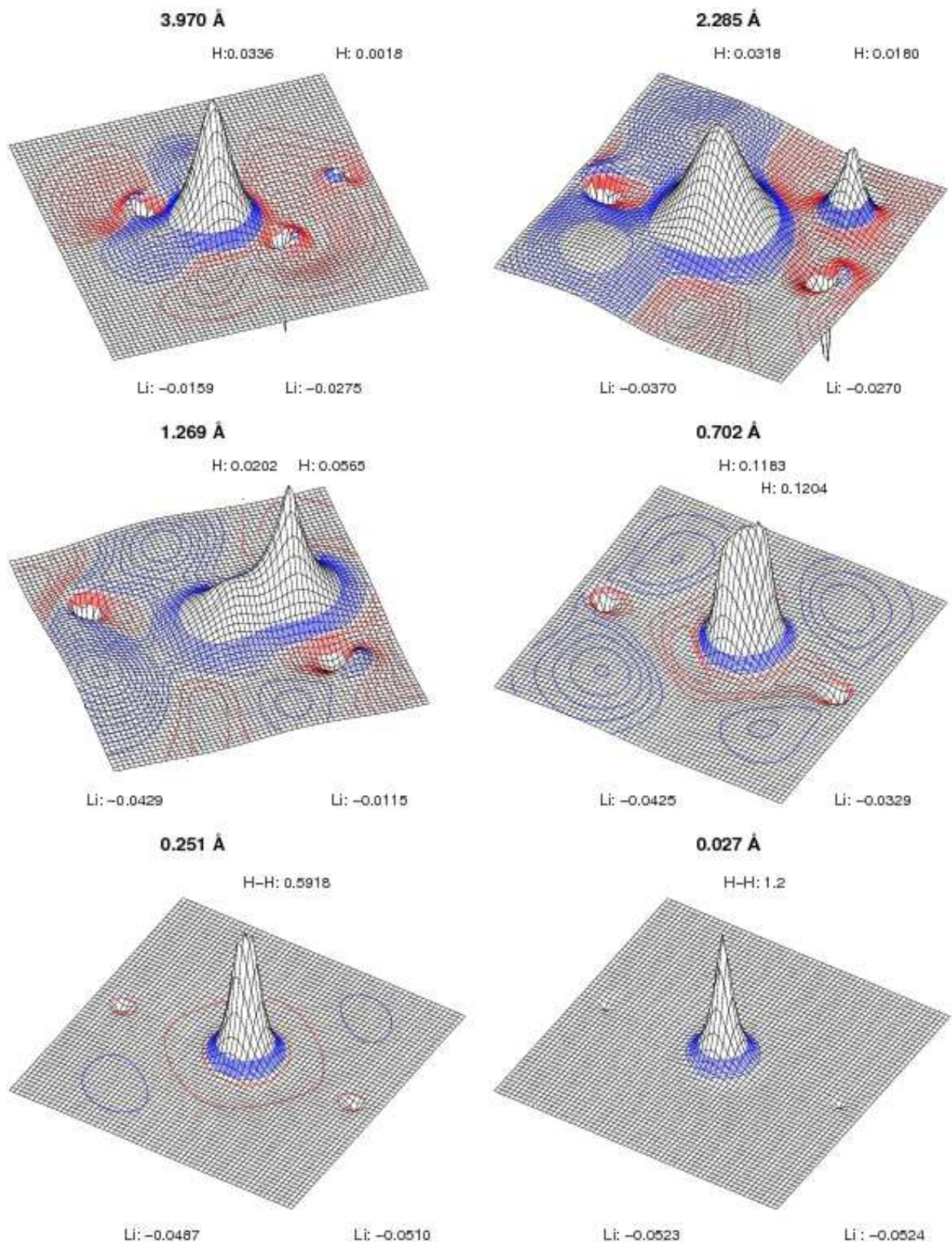}} \par}

\caption{\label{fig:MomentBDLil} Instant recordings of the bond density during the
impact in the molecule \texttt{lih-li1} with the basis 6-31G(d,p)}
\end{figure*}

\begin{figure}
{\par\centering \resizebox*{1\columnwidth}{!}{\includegraphics{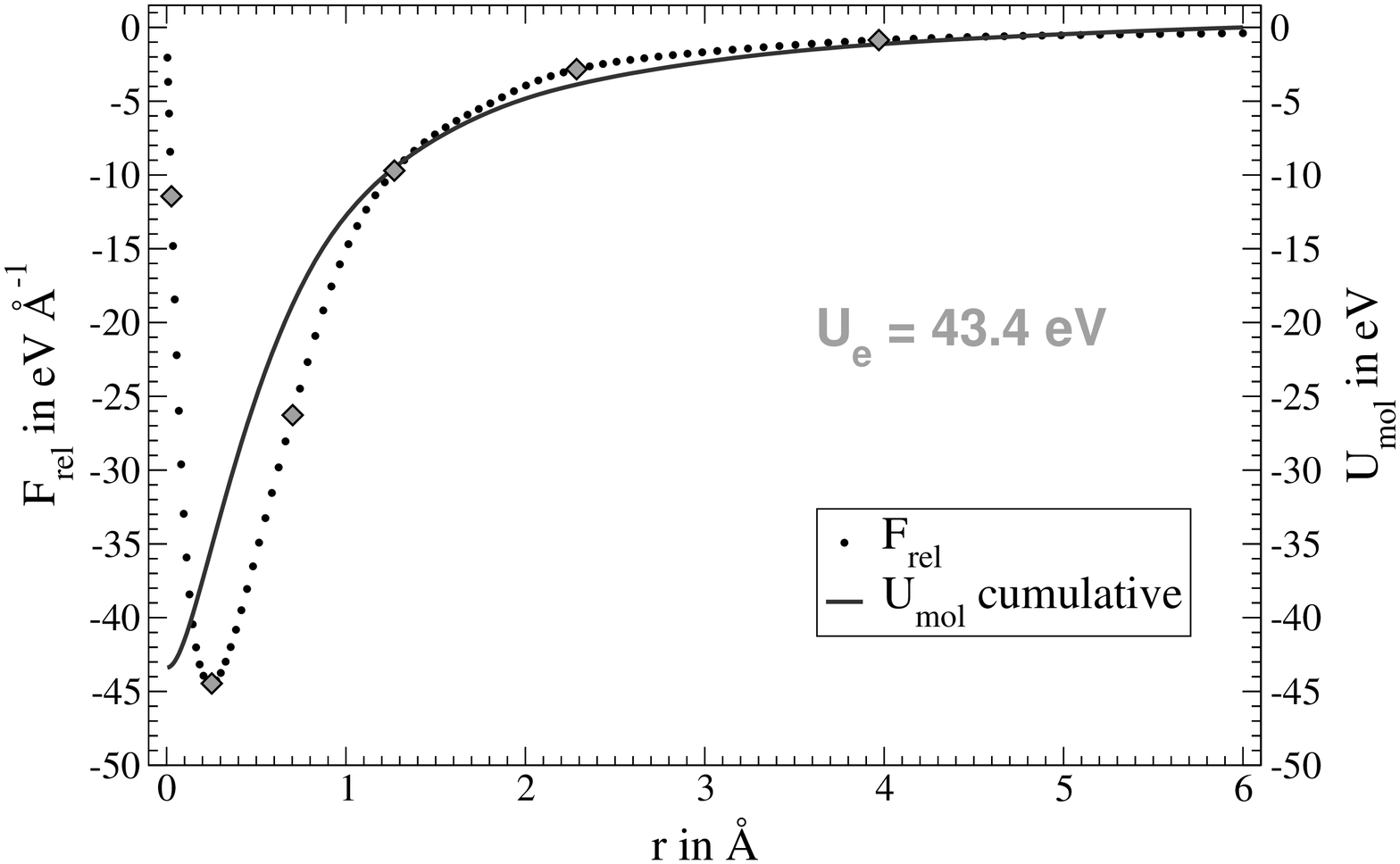}} \par}

\caption{\label{fig:RelF-Lil631}Relative force and molecular screening energy at the
impact in the molecule \texttt{lih-li1} with the basis 6-31G(d,p).}
\end{figure}

During the quasi static simulation of the impact process instant recordings
of the bond density at various distances between the projectile and the target
were taken which are assorted in fig.~\ref{fig:MomentBDLil}. The plots of the
bond density lie in the intersection of the trajectory of the projectile which
is drawn inclinated in fig.~\ref{fig:Mol.lil-li1} and contains two Li atoms. 
  For a contrast amplification the bond densities are not rendered with equal
scale. At first the bond density is distributed at the projectile and the target.
At a distance of \( 1.269\, \angstrom  \) an even enhanced transfer from the
target to the projectile and the environment can be observed. From there on
the electron transfer from the Li atoms to the hydrogen nuclei approaching each
other increases evermore. The calculations were performed for 100 points on
the trajectory. The relative forces at these points are plotted in fig.~\ref{fig:RelF-Lil631}. 
 The six points of the bond density plots (fig.~\ref{fig:MomentBDLil}) are
highlighted by rhombuses and mark interesting locations. Beginning at a distance
of \( 6\, \angstrom  \) the absolute value of the Coulomb-adjusted relative
force raises up to a maximum value at \( \sim 0.25\, \angstrom  \) and declines
till zero for further decreasing distances which is understandable because the
approaching nuclei virtually form a single one in comparison to the wavelength
of the electrons. In so far the behaviour is in accord with the TDHF calculations
in \cite[fig. 8]{shoppa96} except for the oscillations due to the time dependence.
The full curve in fig.~\ref{fig:RelF-Lil631} is the molecular screening energy
which was calculated according to (\ref{m:screen.PfadInt}) along the trajectory
at every point. The final result is with \( -43.39\, \electronvolt  \) way
too low compared to the experimentally obtained values (\( \lesssim 150\, \electronvolt  \)
for Li\footnote{%
The results for the highly reactive alkaline metals Li and Na are heavily impaired
by the surface oxidation under ion irradiation \cite{nimb06}.
} \cite{prc07} and \( [190,320]\, \electronvolt  \) for the other metals \cite{volos98,europhys01,dis}).

\begin{figure}
{\par\centering \resizebox*{1\columnwidth}{!}{\includegraphics{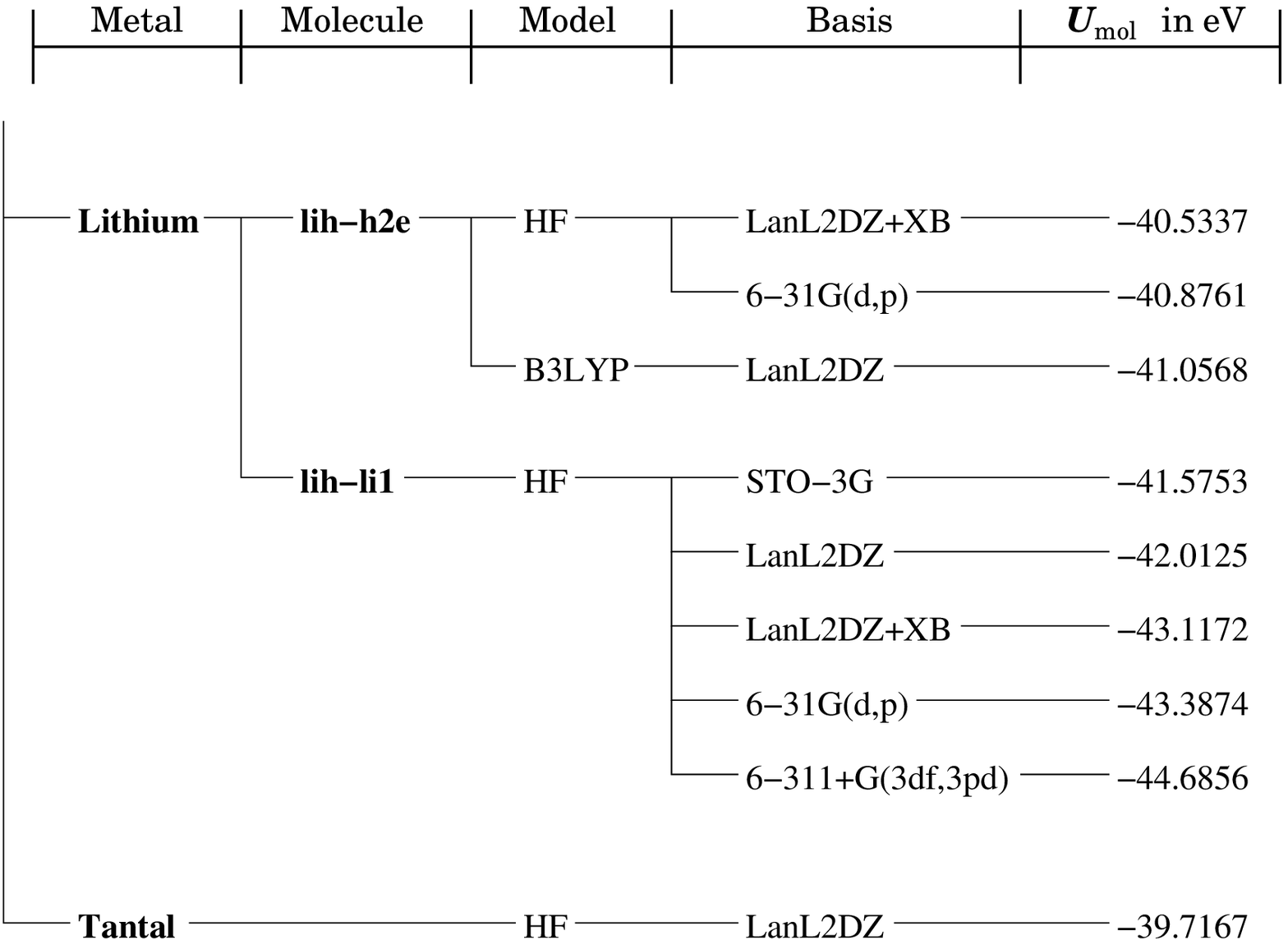}} \par}

\caption{\label{fig:g94result} Overview of the calculation results.}
\end{figure}

The reasons for this unsatisfying result are to be sought in the hierarchy of
model assumptions, approximations, and simplifications. Only very few of them
could be remedied within the scope of this work. However, some experiments with
the bases could be employed whose restrictive impact on the movability of the
electrons particulary in the conduction band is of jutting importance. The results
of these calculations are summarized in the overview in fig.~\ref{fig:g94result}. 
 They are arranged according to increasing complexity of the bases, in each
case. Though from the smallest to the largest basis the value of \( U_{\mathrm{mol}} \)
becomes larger only by \( 10\% \). Yet the tendency is clear: the larger the
basis the higher the screening potential. With an extension of any kind no physical
error can be committed since the electrons arrange due to the energy minimization.
Only the restriction of the bases induces errors. Because the established available
bases are centered at the atoms they are hardly suitable for the simplified
description of the delocalized electrons of the conduction band.

\begin{figure}
{\par\centering \resizebox*{1\columnwidth}{!}{\includegraphics{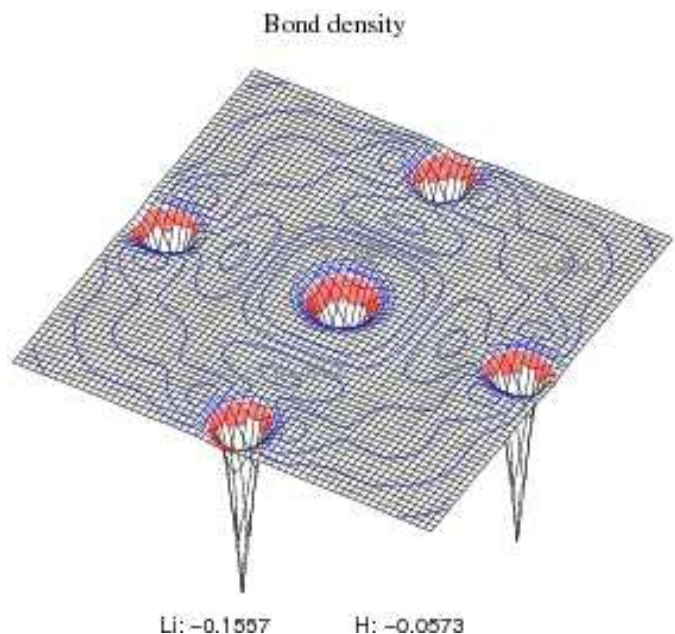}} \par}

\caption{\label{fig:BindDLil}Bond density in the molecule \texttt{lih-li1} with the
basis STO-3G.}
\end{figure}

The effect of the bases on the bond density of the molecule can be assessed
by comparison with the smallest basis STO-3G whose bond density is depicted
in fig.~\ref{fig:BindDLil}. 
 Unlike the 6-31G(d,p) basis in fig.~\ref{fig:DichteLil}.b no electron transfer
to the hydrogen can take place here because the too small basis does not permit
enough mobility for the electrons. Instead only between the nuclear positions
a weak conglomeration accrues like for the covalent bond.

\begin{figure}
{\par\centering \resizebox*{1\columnwidth}{!}{\includegraphics{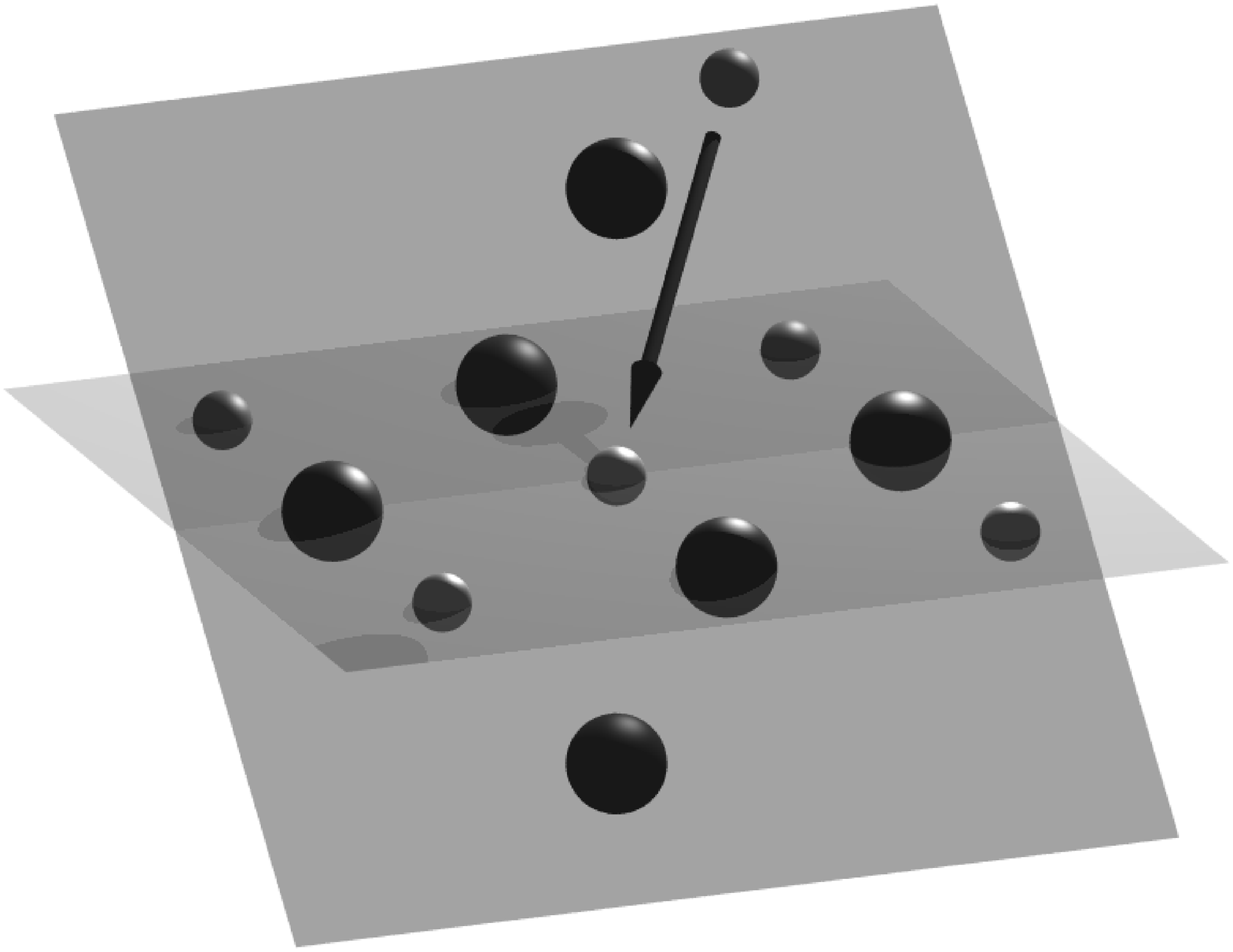}} \par}

\caption{\label{fig:Mol.lih-h2e}LiH molecule (\texttt{lih-h2e}).}
\end{figure}

\begin{figure*}
{\par\centering \resizebox*{1\textwidth}{!}{\includegraphics{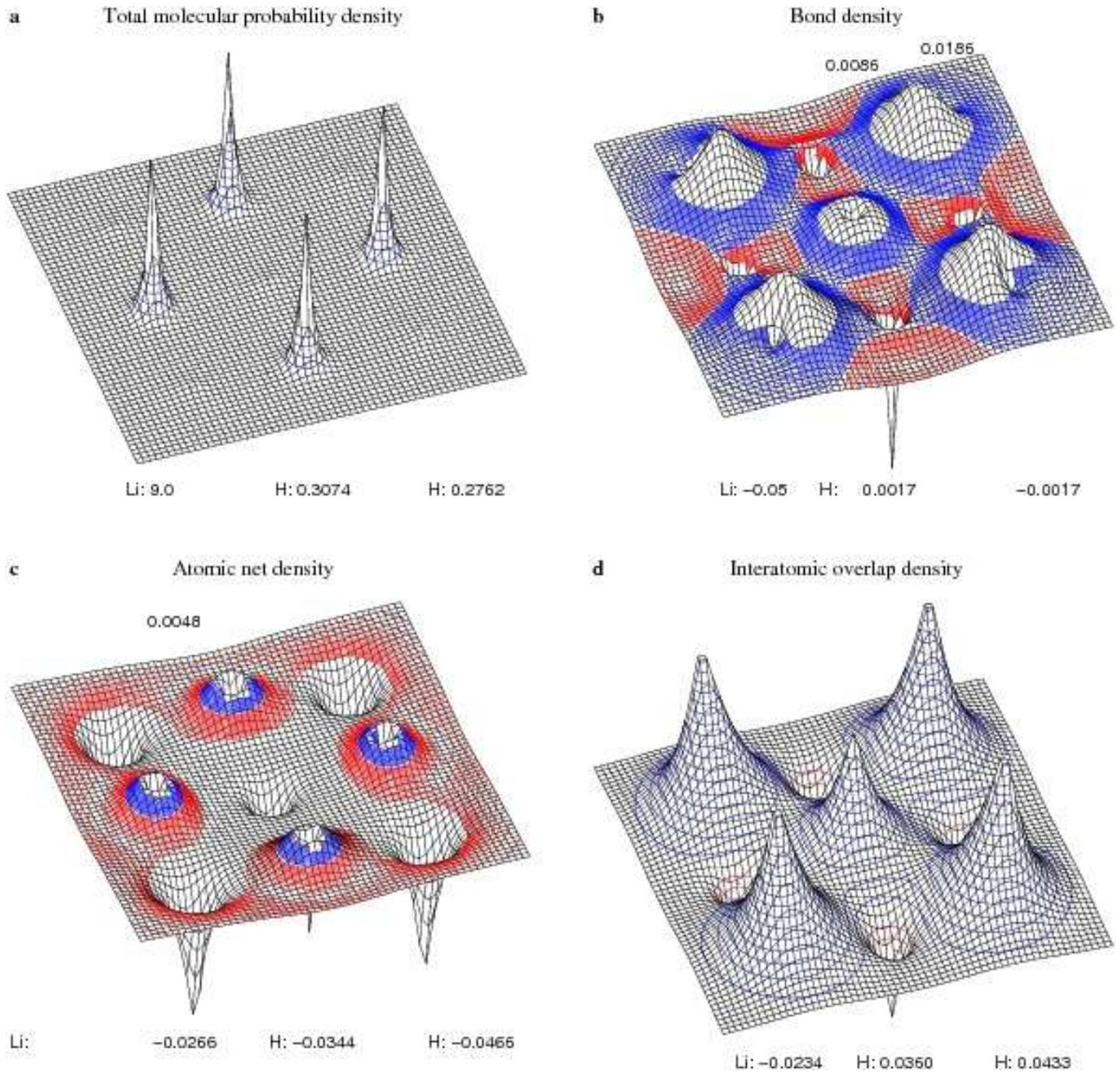}} \par}

\caption{\label{fig:DichteLiH}Different electronic densities in the molecule \texttt{lih-h2e}
with the basis 6-31G(d,p)}
\end{figure*}

The impact of the size of the molecule can only be estimated with narrow limitations,
since by the addition of further spheres of H and Li atoms around the central
target the problem would not be feasible for the workstation any more. Hence,
only on one plane further H atoms were added as fig.~\ref{fig:Mol.lih-h2e}
shows. 
 The belonging density plots in the plane of the Li atoms (fig.~\ref{fig:DichteLiH})
differ perceptibly from those of the simpler molecule (fig.~\ref{fig:DichteLil}). 
 From the comparison of the bond densities, the atomic net and the overlap densities
it is apparent that the perimeter H atoms attract electrons on the expense of
the central H atom and thus have a higher influx. Accordingly the screening
potential is also lower (fig.~\ref{fig:g94result}). This behaviour is vitally
due to the circumstance that the perimeter H atoms have no Li partners which
supply their electrons. If those Li atoms were present the transfer of electrons
from the central H atom would be quite perspicuously lower. It is interesting
that the maximum values of the bond density form a ring around the central H
atom and the maximum values of the perimeter H atoms are not centered around
them but shifted towards the central H atom.

Indeed the more complex basis brings higher values for the screening potential
here again. Furthermore, it is here also documented how a method affects going
beyond the Hartree-Fock method videlicet the density functional theory with
the three parameter functional of Becke, B3LYP \cite{becke88,becke88b,lee88}.
The more accurate provision for the electron correlation effects causes a stronger
repulsion of electrons with equal spins and enhances the screening potential.
However, this effect is not very large.

A further possibility for a comparison is provided by the Mulliken population
analysis which assigns the atoms of the compound a charge according to the electron
transfer. Table~\ref{tab:Atomladungen} summarizes those results. 
\begin{table}

\caption{\label{tab:Atomladungen}Atomic charges originating from the Mulliken population
analysis.}
{\centering \begin{tabular}{|c|rrr||r|c|}
\hline 
&
\multicolumn{1}{|c}{\texttt{lih-li1}}&
\multicolumn{1}{c}{\texttt{lih-li1}}&
\multicolumn{1}{c||}{\texttt{lih-h2e}}&
&
\\
&
\multicolumn{1}{|c}{6-31(d,p)}&
\multicolumn{1}{c}{STO-3G}&
\multicolumn{1}{c||}{6-31(d,p)}&
\multicolumn{1}{|c|}{LanL2DZ}&
\\
\hline 
H&
-0.455479&
-0.054858&
-0.255553&
-0.382043&
H\\
Li&
0.336273&
0.108454&
0.126207&
0.030356&
Ta\\
Li&
-0.054266&
-0.040512&
0.126207&
0.030355&
Ta\\
Li&
-0.054267&
-0.040512&
0.187523&
0.145028&
Ta\\
Li&
0.336273&
0.108454&
0.126207&
0.015638&
Ta\\
Li&
-0.054266&
-0.040512&
0.126207&
0.145028&
Ta\\
Li&
-0.054267&
-0.040512&
0.187523&
0.015637&
Ta\\
H&
&
&
-0.156081&
&
\\
H&
&
&
-0.156081&
&
\\
H&
&
&
-0.156081&
&
\\
H&
&
&
-0.156081&
&
\\
\hline 
\end{tabular}\par}\end{table}
 The topmost atom is the central target H atom followed by the innermost sphere
of metal atoms and further perimeter H atoms. The charges are specified in multiples
of the elementary charge. The larger basis for Li provokes an almost ten times
higher charge transfer to the H atom compared to the smallest one which yields
almost \( 0.5\, e \) and underlines the ionic character of the bond. The extended
molecule with the four perimeter H atoms permits recognition of the reduction
of the charge at the central H atom and the carry-over to the perimeter H atoms.
In comparison the rather high differences in the charge transfers were not reflected
to this extent in the differences of the screening potentials (fig.~\ref{fig:g94result}).
The reason for it is that the charge assignment to the atoms is not a quantum
mechanical observable and hence tainted with a certain arbitrariness. This can
also be seen at the smallest molecule whose Li atoms are fully symmetrically
arranged where two of them donate and four sparsely ingest electrons.

\begin{figure}
{\par\centering \resizebox*{1\columnwidth}{!}{\includegraphics{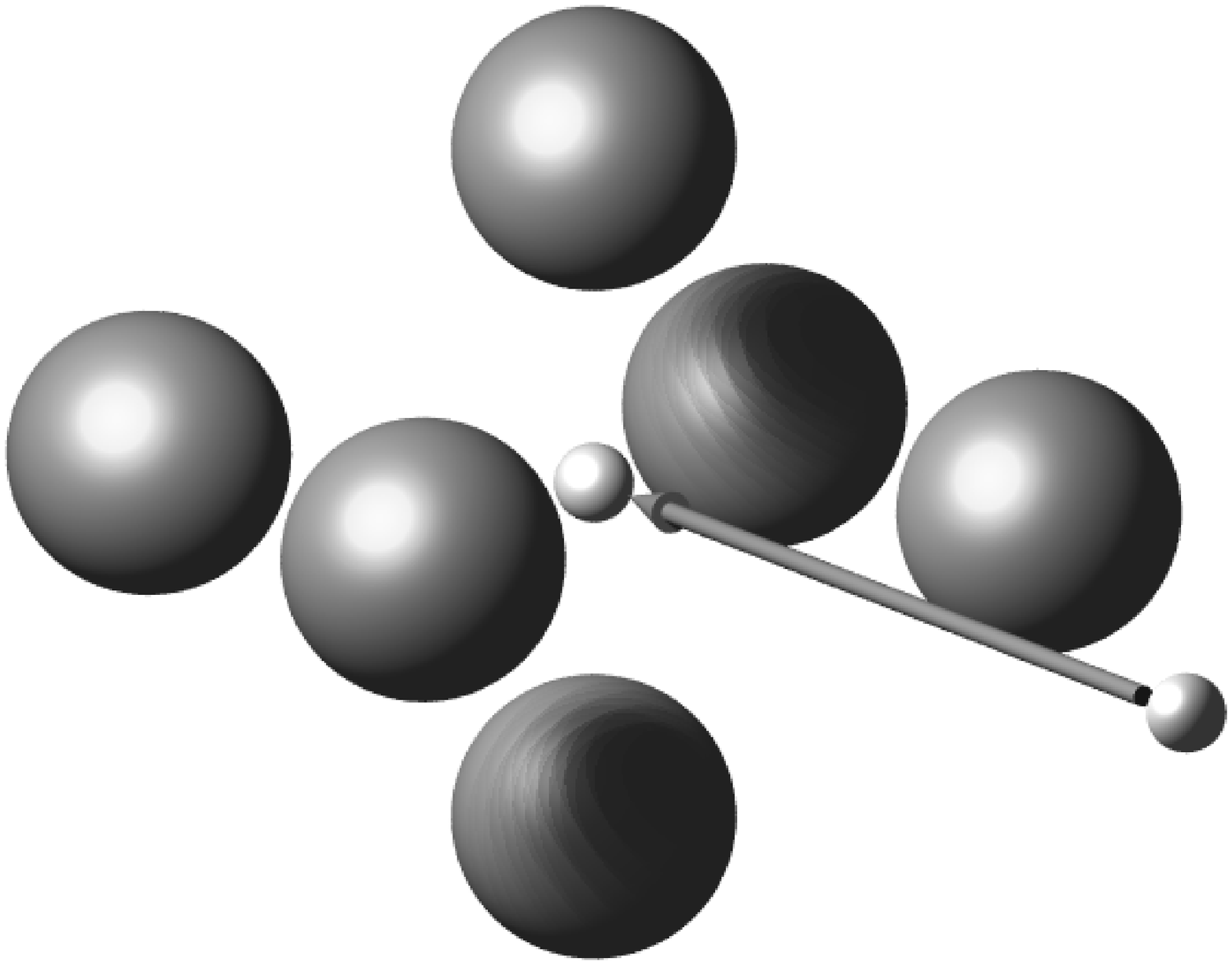}} \par}

\caption{\label{fig:Mol.tah}\protect\( \mathrm{Ta}_{2}\mathrm{H}\protect \) molecule
with projectile trajectory.}
\end{figure}

\begin{figure}
{\par\centering \resizebox*{1\columnwidth}{!}{\includegraphics{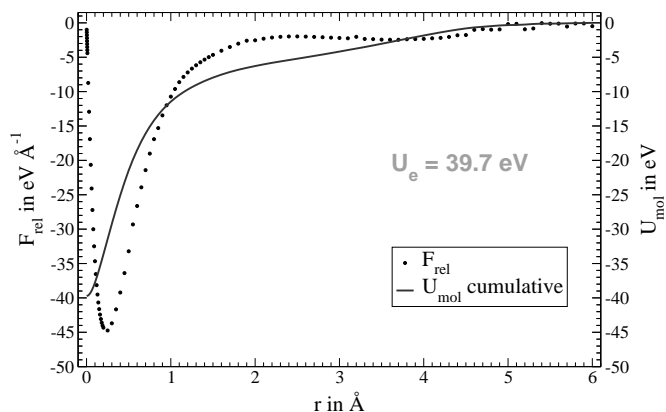}} \par}

\caption{\label{fig:RelF-Tat}Relative force and molecular screening energy at the impact
in the molecule \protect\( \mathrm{Ta}_{2}\mathrm{H}\protect \).}
\end{figure}

In the right part of the table charge assignments for Tantalum are listed. Owing
to the considerable computer time consumption by virtue of the many electrons
to be included only one simulation run on the smallest molecule with the simpliest
basis could be performed. The result of \( U_{e}=39.72\, \electronvolt  \)
even stays behind the smallest Li value, however the charge assignment to the
central H atom is not much smaller. The molecule and the projectile trajectory
are shown in fig.~\ref{fig:Mol.tah}. 
 The behaviour of the relative force in fig.~\ref{fig:RelF-Tat} is remarkable. 
  The oscillations in the far field were caused by the large electron clouds
with high angular momentum of the Ta atoms.

For details refer to \cite{dis}.

\section{Conclusion}

The calculated values for the electron screening energies stay clearly below
the experimental ones. The simplifications in the numerical model were enforced
by the available computing power. The performed numerical simulations consumed
already a netto computation time of \( 1.6\, \mathrm{a} \). Single points on
the trajectory took one week. Anyhow, the simulations could show the migration
of electrons from the host metal atoms to the hydrogen both for the target nuclei
and the projectile in the impact. This migration becomes larger when the number
of base functions is increased. However, in the electron transfer only the immediate
adjacent atoms are involved and no distant atoms. This is due to the limited
size of the basis which are designed for molecules and not for larger solids.
The further extension of the basis sets would provide more freedom of mobility
for the electrons. Particularly, the modelling of conduction band states as
provided by Bloch wave functions could be an asset. But according to the computational
cost function already this extensions to the quasi-static model with an accompanying
increase of the crystal size would require the next scale of computational power
as computing clusters or super computers. Furthermore, time dependent effects
can be expected to contribute the larger share in attaining the experimental
screening results particularly when collective interactions between the collision
partners and many electrons from a larger region are to be included like plasmons,
convoy electrons and effects of the Fermi surface. But a prerequisite for it
are a high electron mobility provided by large basis sets. This would require
the complete redesign of a computer code for a massive parallel computer worth
many man years of work. But then very instructive insights into the electron
dynamics for the understanding of the mechanism can be expected which is not
accessible in the experiment so far. In exchange for higher code development
effort the computational costs could possibly be reduced by three orders of
magnitude using the new \$-operator formalism for the solution of partial differential
equations by \cite{starkl99}.

\end{document}